\documentclass{article}

\usepackage{microtype}
\usepackage{graphicx}
\usepackage{subcaption}
\usepackage{booktabs} 

\usepackage{hyperref}
\usepackage{wrapfig}

\usepackage[preprint]{icml2026}

\usepackage{amsmath}
\usepackage{amssymb}
\usepackage{mathtools}
\usepackage{amsthm}

\usepackage{booktabs}   
\usepackage{multirow}   
\usepackage{graphicx}   
\usepackage[normalem]{ulem}
\usepackage[capitalize,noabbrev]{cleveref}

\theoremstyle{plain}

\theoremstyle{definition}

\theoremstyle{remark}

\usepackage[textsize=tiny]{todonotes}

\icmltitlerunning{Radio-Frequency Inverse Rendering for Wireless Environment Modeling}

\begin{document}

\twocolumn[
  \icmltitle{Radio-Frequency Inverse Rendering for Wireless Environment Modeling}

  \begin{icmlauthorlist}
  \icmlauthor{Fuhai Wang}{yyy,comp}
  \icmlauthor{Zihan Jin}{yyy}
  \icmlauthor{Lehang Wang}{yyy}
  \icmlauthor{Xuehui Dong}{yyy}
  \icmlauthor{Tiebin Mi}{yyy}
  \icmlauthor{Robert Caiming Qiu}{yyy}
  \icmlauthor{Zenan ling}{yyy}
  \end{icmlauthorlist}

  \icmlaffiliation{yyy}{School of Electronic Information and Communications, Huazhong University of Science and Technology, Wuhan, China}
  \icmlaffiliation{comp}{Institute of Artificial Intelligence, Huazhong University of Science and Technology, Wuhan, China}

  \icmlcorrespondingauthor{Zenan ling}{lingzenan@hust.edu.cn}

  \icmlkeywords{Machine Learning, ICML}

  \vskip 0.3in
]



\printAffiliationsAndNotice{}  

\begin{abstract}
Neural rendering paradigms have recently emerged as powerful tools for radio frequency (RF). However, by entangling RF sources with scene geometry and material properties, existing approaches limit downstream manipulation of scene geometry, wireless system configuration, and RF reasoning. To address this, we propose a physically grounded RF inverse rendering (RFIR) framework that explicitly decouples RF emission, geometry, and material electromagnetic properties. Our key insight is an RF-aware bidirectional scattering distribution function, embedded into the Gaussian splatting paradigm as an RF rendering equation. Each Gaussian primitive is endowed with intrinsic physical attributes, including surface normals, material electromagnetic parameters, and roughness, and leveraged by a customized ray-tracing scheme to represent RF signal synthesis. The proposed RFIR generalizes three typical RF tasks: radar cross-section synthesis, received signal strength indicator prediction, and wireless scene editability. Experiments demonstrate significant performance advantages, underscoring the potential for wireless world modeling.
\end{abstract}

\section{Introduction} \label{Sec:Introduction} 
Recently, a growing body of work~\cite{zhao2023nerf2, lu2024newrf, wen2024neural, wen2025wrf, yang2025gwrf, yang2024r, hoydis2024learning} has actively explored neural volumetric rendering in radio-frequency systems, leveraging neural radiance fields (NeRF) or 3D Gaussian splatting (3DGS) to reconstruct a continuous radio radiance field (RRF) directly from wireless measurements. These approaches not only model the underlying physical propagation mechanisms of radio-frequency signals, but also leverage the strong spatial representation capability of neural radiance fields, enabling broad applicability to downstream wireless tasks such as wireless digital twins~\cite{jiang2025learnable, pharr2023physically}, integrated sensing and communication~\cite{Wei2025Integrated}, and intelligent coverage optimization~\cite{10465179}.


Existing approaches~\cite{zhao2023nerf2, lu2024newrf, wen2024neural, wen2025wrf, yang2025scalable, yang2025gwrf, yang2024r} use voxels or 3D Gaussian primitives to directly parameterize radiated RF signals, explicitly modeling both amplitude and phase. While expressive, this design fundamentally entangles signal emission with scene geometry and material properties~\cite{zhao2023nerf2, lu2024newrf, wen2024neural}. Such tight coupling undermines scene generalization: even minor geometric or material changes typically invalidate the learned representation and require costly retraining. This limitation substantially constrains their applications to several practically important downstream tasks that require decoupling of signal sources from  propagation channels, such as RF attribute representation of the environment~\cite{chen2024rfcanvas,Liu2026SceneSB}, wireless system reconfiguration via transmitter (Tx) and receiver (Rx) repositioning~\cite{wang2025dreamer,wang2025source,zhao2023nerf2}, and RF scene editability, i.e., predicting signal behavior under hypothetical scene or system modifications~\cite{zhu2024physics}. 

To bridge this gap, we propose an RF inverse rendering (RFIR) framework that decomposes observed RF signals into scene geometry, material RF properties, and signal emission. Unlike prior approaches that directly predict per-point signal emission, we treat each Gaussian as a local channel response between input and output signals. We further perform fine-grained synthesis of global wireless signals by modeling free-space path loss, visibility through ray tracing, and aggregation of Gaussian-scattered fields via alpha blending. As such, RFIR decouples the signal source from the propagation environment, enabling flexible, generalizable, and physically grounded RF scene representations.

Technically, RFIR integrates the physical fidelity of an RF-aware bidirectional scattering distribution function (RF-BSDF) into the 3DGS framework to explicitly capture wireless signal-environment interactions. Each Gaussian primitive is endowed with intrinsic, learnable attributes, including surface normals, material electromagnetic parameters, and surface roughness, while visual priors are leveraged to ensure geometric consistency in normal estimation. We further develop a customized CUDA-based ray tracing module operating directly on Gaussian primitives, which explicitly computes incident-field visibility and RF propagation paths, enabling efficient RF signal synthesis and inverse rendering within a unified, physics-aware framework.

The proposed decomposition framework naturally generalizes to three representative downstream tasks, including Wi-Fi radar cross section (RCS) synthesis at 2.4/5.8 GHz and wideband settings, received signal strength indicator (RSSI) prediction, and wireless scene reconfigurability. Extensive experiments on measured and simulated data show that our approach consistently outperforms state-of-the-art (SOTA) RF modeling methods, demonstrating strong generalization across diverse and practical applicability.

In summary, our key contributions are  as follows:
\begin{itemize}
\item We propose RFIR, a decoupled framework that decomposes observed signals into scene geometry, material RF properties, and signal emission. It explicitly models free-space path loss, geometry-dependent visibility via ray tracing, and aggregation of Gaussian-scattered fields, enabling flexible RF scene representations.
\item We integrate an RF-BSDF into the 3DGS framework to capture signal–environment interactions. Each Gaussian has learnable attributes and a specialized CUDA-based ray tracing module computes visibility and propagation paths, allowing efficient RF signal synthesis.
\item Extensive evaluations on measured and simulated data for RCS synthesis, RSSI prediction, and RF scene editability, demonstrate that our method consistently outperforms advanced baselines. 
\end{itemize}

\begin{figure*}[t]
    \centering
    \includegraphics[width=1\linewidth]{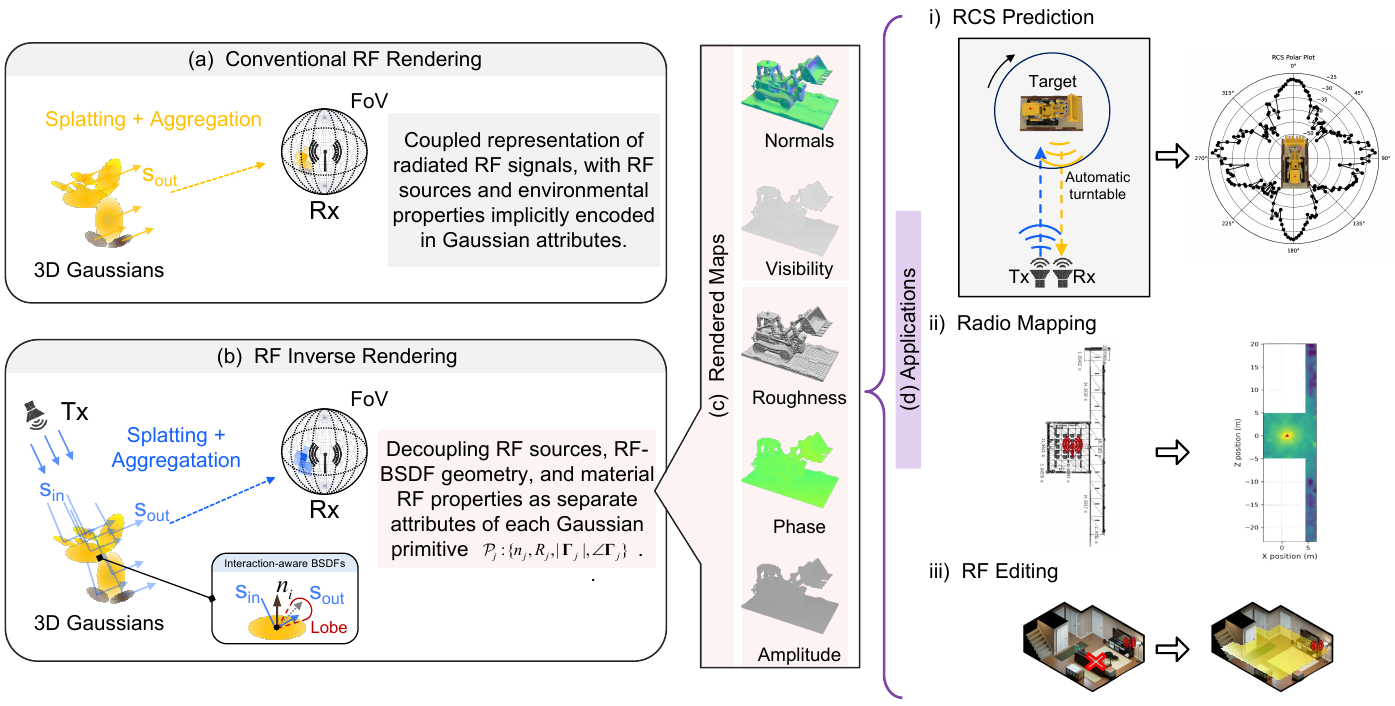}
    \caption{RF Inverse Rendering. (a) Classical RF rendering and (b) the proposed RF inverse rendering. Inverse rendering enables decoupled RF attribute rendering (c) and broader applications (d). }
    \label{fig:Fig1_comparison_and_tasks}
\end{figure*}

\begin{figure*}[t]
    \centering
    \includegraphics[width=0.98\linewidth]{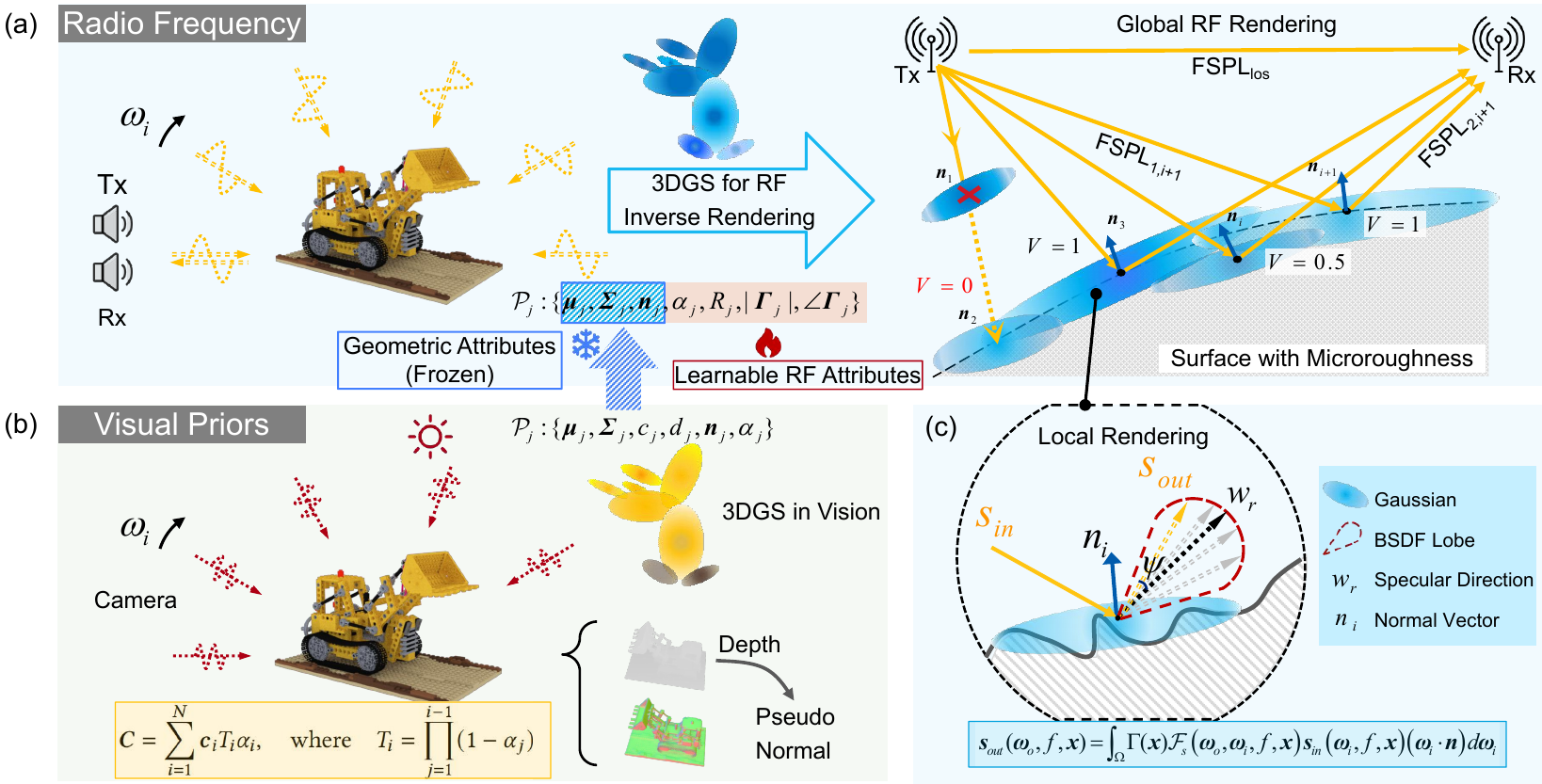}
    \caption{Overview of system workflow}
    \label{fig:pipeline}
\end{figure*}

\section{Related Work} \label{Sec:Related Work}
\paragraph{Neural-based Channel Modeling.} Neural-based channel modeling methods broadly fall into two categories: neural-enhanced ray tracing (RT) and neural volumetric RF rendering. To approximate material RF properties, neural-enhanced RT integrates learning-based components into classical RT pipelines, such as learning complex reflection coefficients consistent with Snell’s law~\cite{han2025rayloc, jiang2025learnable} or modeling scattering patterns with multilayer perceptrons (MLPs)~\cite{orekondy2023winert}. However, these methods~\cite{zhu2024physics,chen2024rfcanvas} critically rely on high-quality 3D mesh representations, which are costly to acquire in practice, and their simplified physical assumptions often fail to capture complex scattering effects, leading to a notable simulation-to-reality gap. 

Neural volumetric RF rendering methods~\cite{zhao2023nerf2, wen2024neural, yang2025gwrf, yang2025scalable, yang2024r} directly parameterize radiated RF signals with voxels or 3D Gaussian primitives, explicitly modeling amplitude and phase under fixed or weakly parameterized wireless settings. The signal source, propagation channel, and scene geometry are tightly coupled within a single neural representation, limiting flexibility when changing the wireless configuration. As a result, these methods primarily output radiated signals, rather than modeling the underlying channel responses that govern signal propagation and interaction with the environment.

In contrast, our method explicitly decouples signal sources from environment-dependent responses and incorporates geometry-aware visual priors, enabling more robust generalization and higher-fidelity RF inverse rendering across diverse wireless scenarios, as shown in Fig.~\ref{fig:Fig1_comparison_and_tasks}. 

\paragraph{Inverse Rendering.} Inverse rendering in the optical domain typically aims to decompose observed appearance into scene geometry, material properties, and illumination~\cite{gao2024relightable,zhu2024gs,liang2024gs,shi2025gir,yao2022neilf,zhang2021physg,zhang2021nerfactor}. While effective for image synthesis, this design limits their applicability to downstream tasks such as relighting and scene editing. 

In this work, we study inverse rendering for modeling the wireless world. General inverse rendering methods~\cite{chen2018computational, deshmukh2022physics, gu2025irgs} are fundamentally derived from Maxwell's equations under wave-optics formulations. These approaches operate on complex-valued field measurements, and often employ integral-equation formulations with explicit regularization. Their objective is to recover dielectric properties at optical frequencies, object geometries, and other physical parameters~\cite{han2025rayloc, hehn2024differentiable}. 

\paragraph{BSDF for RF Modeling.} The BSDF~\cite{zhang2025Unified} characterizes the interaction between EM waves and objects by modeling the input-output relationship of incident and outgoing fields. In optics~\cite{walter2007microfacet, Xu2009Bidirectional, Wei2025Integrated}, BSDFs are widely used to represent diverse light-matter interactions. By adjusting BSDF parameters, various phenomena such as diffuse reflection and specular reflection can be accurately modeled, enabling photorealistic rendering across different spectral bands in computer graphics~\cite{pharr2023physically}.

When extending BSDF-based modeling from optics to the RF domain, electromagnetic coherence effects become more pronounced, with wave behavior dominating over particle-like approximations~\cite{gu2025irgs, steinberg2024free}. As a result, BSDF models must explicitly account for both amplitude and phase responses across frequency bands. While preliminary measurements have been explored in RCS settings~\cite{vitucci2023reciprocal, miao2018modeling, degli2007measurement}, the application of RF-BSDF models within realistic wireless system scenarios remains limited. 

\section{Preliminaries}
This section reviews NeRF and 3DGS as neural volumetric representations for wireless channel modeling.
\paragraph{NeRF-based Channel Modeling.}
To model radio radiance fields, NeRF-based methods~\cite{zhao2023nerf2, luo2025fermi, amballa2025can} employ implicit volumetric representations and use a MLP to predict the radiance $\boldsymbol{s}_i$ at each sampled point. The effects of the signal source, scene geometry, and material properties are tightly coupled and implicitly absorbed into voxel-wise radiance predictions, resulting in an entangled representation of the RRF.

The effective opacity $\alpha_i$ is also derived from the predicted volume density $\sigma_i$, produced by an attenuation network, together with the sampling interval $\delta_i$, as $\alpha_i = 1 - \exp(-\sigma_i \delta_i)$, modeling the interaction probability between the propagating field and each volumetric element. Following volumetric radiative transfer, the synthesized signal $\boldsymbol{S}$ along a ray is obtained via $\alpha$-blending:
\begin{equation}
\label{Eq:transmitance_1}
\boldsymbol{S} = \sum_{i=1}^{N} \boldsymbol{s}_i \, T_i \, \alpha_i, \quad
T_i = \prod_{j=1}^{i-1}(1 - \alpha_j),
\end{equation}
where $T_i$ denotes the accumulated transmittance from preceding samples. RF signal rendering is jointly governed by a complex-valued radiance network that predicts the radiated signal $\boldsymbol{s}_i$ and an attenuation network that models propagation-induced channel attenuation, together enabling neural rendering of wireless signal propagation.

\paragraph{3DGS-based Channel Modeling.}
To enable efficient and fast reconstruction of 3D RF fields, 3DGS models the environment using a set of anisotropic Gaussian ellipsoids, each parameterized as an ellipsoidal volumetric distribution~\cite{kerbl20233d, gao2024relightable}. Specifically, the $k$-th Gaussian is parameterized by its mean position $\boldsymbol{\mu}_k \in \mathbb{R}^3$ and a symmetric positive-definite covariance matrix $\boldsymbol{\Sigma} \in \mathbb{R}^{3\times3}$. Its spatial influence is formally defined by the multivariate Gaussian density function:
\begin{equation*} \label{Eq:Gaussians_1}
\mathcal{G}(\boldsymbol{x}|\boldsymbol{\mu}_k,\boldsymbol{\Sigma}) = \exp\left(-\frac{1}{2}(\boldsymbol{x}-\boldsymbol{\mu}_k)^{\mathrm{T}}\boldsymbol{\Sigma}^{-1}(\boldsymbol{x}-\boldsymbol{\mu}_k)\right).
\end{equation*} 
Beyond geometric parameters, each Gaussian primitive is assigned intrinsic appearance attributes: an opacity RF coefficient $\alpha_i \in [0,1]$ and a view-dependent radiance $\boldsymbol{s}_i$, the latter of which is typically parameterized via a set of spherical harmonic coefficients. Attaching RF signals (amplitude and phase) to Gaussian primitives is analogous to representing RGB channels in images, such that both RF and classical visual geometry reconstruction treat Gaussians as radiative entities, following a consistent modeling formulation.

\paragraph{Estimation of Gaussian Normal.}
The surface normal $\boldsymbol{n}$ for the scene constructed from Gaussians can in principle be estimated from either RF signals or visual information~\cite{zhu2024gs,gao2024relightable,zhang2021physg,shi2025gir}, and the resulting RF- and visually-derived normals are expected to be consistent with each other. Subsequently, the surface normal $\boldsymbol{n}$ and a depth attribute $d$ are incorporated for each Gaussian primitive, and an optimization strategy tailored for RF-aware physically based rendering is devised, as detailed in the Appendix~\ref{App:Geometric Regularization}. Specifically, the hybrid logic in~\eqref{Eq:transmitance_1} is extended to synthesize the scene's depth map $\mathcal{D}$ and normal map $\mathcal{N}$ as follows:
\begin{equation*} \label{Eq:DN}
\{\mathcal{D}, \mathcal{N}\} = \sum_{i \in \mathcal{N}} \alpha_i T_i \{d_i, \mathbf{n}_i\}
\end{equation*} 
where $d_i$ denotes the $z$-depth coordinate of the $i$-th Gaussian in the view space. By utilizing the same accumulated transmittance $T_i$ and effective opacity $\alpha_i$ as in the visual rendering process, we ensure a spatially consistent integration of geometric attributes across the rendered scene. To improve the robustness of visual guidance, we introduce supplementary constraints, detailed in Appendix~\ref{App:Depth Uncertainty}.

\section{RF Inverse Rendering}\label{sec:System Design}
This work aims to develop a physically grounded RF inverse rendering (RFIR) framework that decouples RF emission, scene geometry, and material RF properties for flexible and interpretable wireless scene modeling. Specifically, we embed a parameterized BSDF into the 3DGS framework to support forward RF propagation. We use 3DGS solely to model the geometric structure of scene objects, while several key RF-BSDF attributes, including effective roughness and complex reflection coefficients, are recovered via inverse rendering from Tx-Rx RF signal observations. This physically based rendering formulation decouples the wireless channel from signal source configurations, enabling scene reconfiguration and dynamic editing. An overview of our pipeline is shown in Fig.~\ref{fig:pipeline}. 

To ensure accurate channel characterization, visual-prior-guided geometry initialization is adopted to reconstruct a coarse topological representation of the scene, which serves as the geometric foundation for subsequent 3DGS-based modeling. Details of the visual geometry reconstruction are extensively described in prior works~\cite{gao2024relightable,zhu2024gs,shi2025gir,zhang2021physg,zhang2021nerfactor}. Due to the difficulty of estimating $\boldsymbol{n}$ directly from RF, a visual-based approach~\cite{gao2024relightable} is adopted and applied in advance during the visual 3DGS reconstruction stage. 

\subsection{Gaussian Primitives: Geometric and RF Attributes}
In summary, our framework represents the wireless environment as a collection of 3D RF Gaussians. The $j$-th Gaussian $\mathcal{P}_j$ is fully parameterized by its geometric and RF attributes: 
\[\{ \boldsymbol{\mu}_j, \boldsymbol{\Sigma}_j, \boldsymbol{n}_j, \alpha_j, R_j, |\boldsymbol{\Gamma}_{j}|, \angle \boldsymbol{\Gamma}_{j} \},\] 
where $\boldsymbol{\mu}_j$ and $\boldsymbol{\Sigma}_j$ characterize the position and shape of the Gaussian, $\boldsymbol{n}_j$ represents the Gaussian normal, $\alpha_j$ denotes the opacity of the RF signal, $R_j$ corresponds to the effective roughness, and $|\boldsymbol{\Gamma}_{j}|$ and $\angle \boldsymbol{\Gamma}_{j}$ denote the magnitude and phase of the complex reflection coefficient, respectively.

During the geometric reconstruction stage, we optimize a vanilla 3D Gaussian point cloud, augmented with an additional normal vector $\boldsymbol{n}$ for each Gaussian. Each vanilla 3D Gaussian primitive is endowed with basic geometric information $\mathcal{P}_j:\{\boldsymbol{\mu}_j, \boldsymbol{\Sigma}_j, \boldsymbol{n}_j\}$. In the RF rendering stage, we preserve the geometry of 3D Gaussians unchanged, i.e., keep the geometric parameters frozen and focus solely on optimizing RF-aware attributes $\{\alpha_j, R_j, |\boldsymbol{\Gamma}_{j}|, \angle \boldsymbol{\Gamma}_{j}\}$ through inverse rendering of the RF signals.

\subsection{Local RF Rendering on a Single Gaussian}\label{Sec:EM-Aware}
Given a set of locked Gaussian primitives $\mathcal{P}_j: \{\boldsymbol{\mu}_j, \boldsymbol{\Sigma}_j, \boldsymbol{n}_j\}$, RFIR employs a complex-valued tensorial field rendering equation~\cite{walter2007microfacet,gu2025irgs} to model electromagnetic wave interaction with each Gaussian primitive, accounting for modified bidirectional scattering distribution function~\cite{walter2007microfacet,ozdogan2019intelligent,partanen2017interference,yao2022neilf} properties and geometry. The radiance $\boldsymbol{s}_{out}$ of each Gaussian primitive, physically computed using the scattering rendering equation, is given by:
\begin{equation*} \label{Eq:signal_gaussian}
    \begin{aligned} 
    \boldsymbol{s}_{out}(\boldsymbol{\omega}_o, f,\boldsymbol{x}) = \sum_{i=0}^{N_s} {{|\boldsymbol{\Gamma}}|e^{j \angle \boldsymbol{\Gamma}}} & {\mathcal{F}}_s\left(\boldsymbol{\omega}_o,\boldsymbol{\omega}_i,f,\boldsymbol{n},\boldsymbol{x}\right) \\ & \boldsymbol{s}_{in}\left(\boldsymbol{\omega}_i,f,\boldsymbol{x}\right)\left(\boldsymbol{\omega}_i \cdot \boldsymbol{n}\right) \bigtriangleup \boldsymbol{\omega}_i 
    \end{aligned} 
\end{equation*} 
where $\mathcal{F}_s(\boldsymbol{\omega}_o,\boldsymbol{\omega}_i,f,\boldsymbol{n},\boldsymbol{x})$ is the RF-aware bidirectional scattering distribution function modeling coherent scattering between incident and outgoing fields.  
$\boldsymbol{\Gamma} \in \mathbb{C}$ denotes the complex reflection coefficient with magnitude $|\boldsymbol{\Gamma}|$ and phase $\angle \boldsymbol{\Gamma}$, which are learnable attributes of each Gaussian. $\boldsymbol{s}_{in}\left(\boldsymbol{\omega}_i,f,\boldsymbol{x}\right)$ represents the incident RF field arriving from direction $\boldsymbol{\omega}_i$ at frequency $f$. The incident directions and their corresponding fields are precomputed based on visibility, as described in Section~\ref{Sec:RFon3DGS}. $N_s$ is the number of discrete incident directions sampled for the Gaussian. $\boldsymbol{n}$ denotes the Gaussian normal vector and $\bigtriangleup \boldsymbol{\omega}_i$ is the solid angle weight for each sampled direction. 

Compared to previous approaches~\cite{zhao2023nerf2, luo2025fermi, amballa2025can} 
that directly model the radiance field using an MLP, this scattering-based rendering formulation represents RF fields and environmental parameters separately, with learnable attributes embedded in each Gaussian primitive, enabling explicit modeling of the physical interactions underlying RF propagation.

\paragraph{RF Bidirectional Scattering Distribution Function Parameterization.} 
To inherently disentangle the modeling of interactions on rough surfaces, we incorporate an RF-aware bidirectional scattering distribution function (RF-BSDF) associated with each Gaussian primitive. This function characterizes the angular distribution of scattered energy resulting from the interaction of an incident wave with a rough surface, as illustrated in Fig.~\ref{fig:pipeline}(c). Following the directive scattering model~\cite{vitucci2023reciprocal}, the scattering pattern ${\mathcal{F}}$ is defined as:
\begin{equation*}~\label{Eq:F1a}
{\mathcal{F}}(\boldsymbol{\omega}_o,\boldsymbol{\omega}_i,\boldsymbol{n},f) = \sqrt{\cos \theta_o} \left( \frac{1 + \cos \psi}{2} \right)^{R} \quad (R \ge 1).
\end{equation*}
Here, $\boldsymbol{n}$ denotes the surface normal, and $f$ is the operational frequency. 
$\boldsymbol{\omega}_i = (\theta_i, \phi_i)$ and $\boldsymbol{\omega}_o = (\theta_o, \phi_o)$ represent the incident and outgoing scattering directions, respectively, with $\theta$ and $\phi$ indicating elevation and azimuth angles relative to the intrinsic surface normal $\boldsymbol{n}$. 
$\psi$ is the angular deviation between the scattering direction $\boldsymbol{\omega}_o$ and the ideal specular direction $\boldsymbol{\omega}_r$, given by Snell's law:
\begin{equation*} 
    \boldsymbol{\omega}_r = \boldsymbol{\omega}_i - 2(\boldsymbol{\omega}_i \cdot \boldsymbol{n}) \boldsymbol{n}.
\end{equation*} 
The exponent $R$ is a learnable effective roughness attribute controlling the scattering lobe's concentration.

By explicitly parameterizing these primitives with surface normals, material-dependent reflection coefficients $\boldsymbol{\Gamma}$, and geometric attributes $R$, our framework successfully decouples the excitation source from the environmental scattering characteristics, enabling high-fidelity scene manipulation.



\subsection{Global RF Rendering via Ray Tracing on Gaussians}\label{Sec:RFon3DGS}
In the following sections, we introduce visibility and blending modules to characterize the two distinct forward propagation stages: the radiation from the Tx to the Gaussian primitives, and the subsequent aggregation of Gaussian-scattered fields at the Rx antenna.

\paragraph{Visibility of Incident Field.} 
Traditional NeRF-based methods~\cite{zhao2023nerf2, luo2025fermi, amballa2025can} model the radiance field $\boldsymbol{s}_{out}$ with MLPs taking Tx/Rx positions as input, omitting the computation of the incident fields at each voxel. To pre-compute the physically meaningful incident fields $\boldsymbol{s}_{in}$ on the surface, we propose a modified visibility method to approximate the environmental RF distribution~\cite{gao2024relightable}. Then the visibility $V$ for each Gaussian point via ray tracing with pointed-based bounding volume hierarchy (BVH)~\cite{karras2012maximizing} from Tx to each Gaussian is defined as: $V_i = (1 - \alpha_{i-1})V_{i-1}$, where $\alpha_{i-1}$ is approximated the contribution of this 3D Gaussian to the ray's opacity, computed using BVH.



As shown in Fig.~\ref{fig:pipeline}(a), the incident RF signal $\boldsymbol{s}_{in}(\boldsymbol{\omega}_i, f)$ at the $i$-th Gaussian primitive, situated $d_i$ away from the Tx, is computed in a fine-grained manner using the Friis transmission formula along with the associated phase shift:
\begin{equation*}~\label{Eq:power_incident_signal}
\boldsymbol{s}_{in}(\boldsymbol{\omega}_i, f) = \sqrt{P_T G_T \frac{\mathcal{A}_i }{4 \pi d_i^2}}  \cdot V_i \cdot \mathcal{S}_{Tx}(f) \cdot e^{-j\phi(d_i)},
\end{equation*}
where $P_T$ and $G_T$ denote the transmitted power and antenna gain, respectively, $d_i$ is the distance from the transmitter to the $i$-th Gaussian, $\mathcal{A}_i$ denotes the geometric intercept factor of the 3D Gaussian, with detailed computations provided in Appendix~\ref{App:projectedcrosssection}. $V_i$ is the visibility of the Gaussian along the propagation path, $\mathcal{S}_{Tx}(f)$ represents the transmitted signal at frequency $f$, and $\phi(d_i) = 2 \pi d_i / \lambda$ is the phase shift over the propagation distance $d_i$, with $\lambda$ denoting the wavelength.


\paragraph{Alpha-Blending for Signal Synthesis.} We adapt 3D Gaussian splatting by projecting Gaussians onto a spherical coordinate system centered at the receiver to align volumetric scattering contributions with the antenna's field of view (FoV; see Appendix~\ref{App:Discretization} for discretization details). Unlike the coupled classical alpha-blending~\cite{zhao2023nerf2}, the RF alpha-blending process separately models free-space path loss, RF-BSDF scattering, and phase shift accumulated from the Tx to the Gaussian primitive $l$. The signal $\mathcal{S}(\omega, f)$ is given by the differentiable volume rendering equation:
\begin{align*}
   &\mathcal{S}(\omega,f)\\
   &= \sum_{l=1}^{N} \boldsymbol{s}_{out}(l;\omega,f) \alpha_{l} \prod_{m=1}^{l-1} \left(1 - \alpha_{m}\right)  \sqrt{ \frac{1}{4 \pi d_m^2} } e^{-j\phi(d_m)} 
\end{align*} 
where $N$ is the number of Gaussians along the ray, $\alpha_l$ is the opacity of the $l$-th Gaussian, and the term $\prod_{m=1}^{l-1} \left(1 - \alpha_{m}\right)$ represents the accumulated transmittance $T_l$ through the preceding Gaussians. The last two terms capture the distance-dependent attenuation and phase shift $\phi$ induced by the propagation distance $d_m$.

The total complex signal $\mathcal{S}(f)$ measured at the receiver is obtained by coherently integrating the volumetric signal contributions $\mathcal{S}(\omega, f)$ across the entire receiver antenna's FoV. The final received complex signal is expressed as:
\begin{equation*} \label{eq:final_csi}
\mathcal{S}(f) = \sum_{i=1}^{N_\omega} \mathcal{C}^R (\omega_i,f) \mathcal{S}(\omega_i,f)
\end{equation*}
where $\mathcal{C}^R (\omega_i,f)$ represents the receiver antenna's pattern, and $N_\omega$ is the total number of sampled discrete directions within the antenna's FoV.

\begin{figure}
    \centering
    \includegraphics[width=1\linewidth]{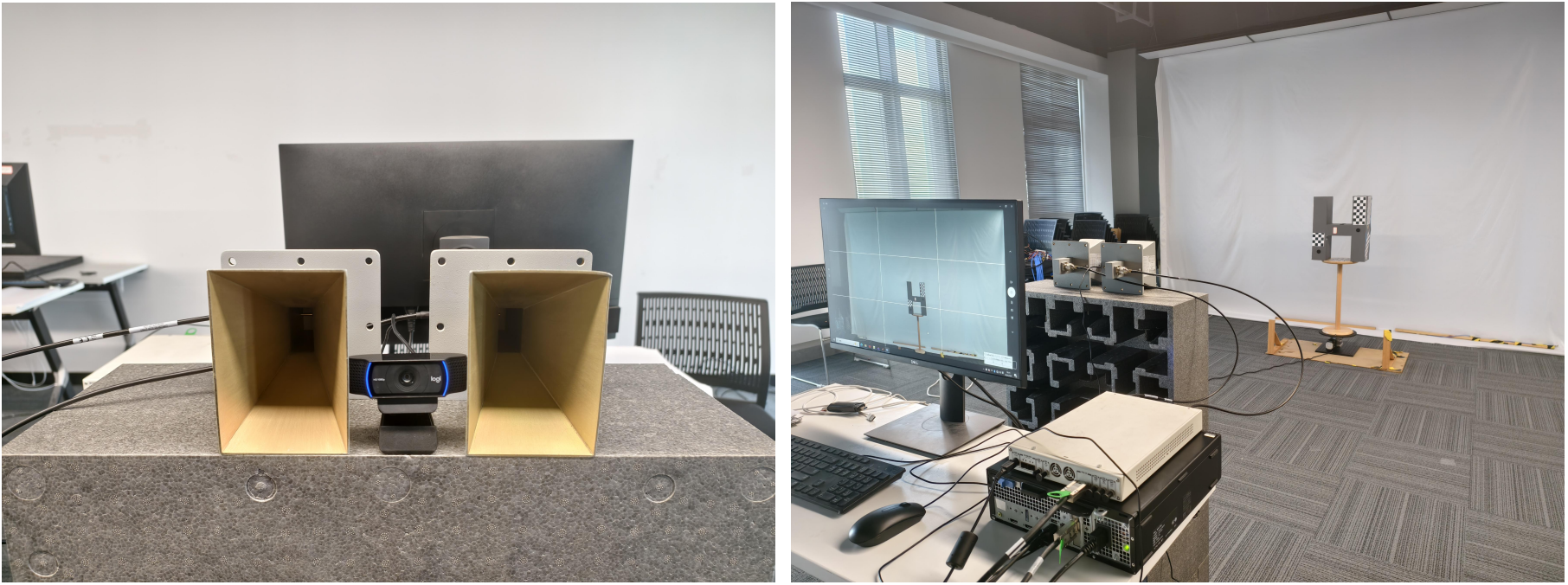}
    \caption{Experimental setup for broadband LFM signal acquisition. Two horn antennas are positioned 3 m from the rotating turntable to transmit and receive the LFM signals, while a camera simultaneously captures multi-view images of the target.}
    \label{fig:LFM_setup}
\end{figure}

\subsection{Loss Function and Algorithm} We design loss functions for RF inverse rendering. Let $\mathcal{S}$ and $RSSI_{Rx}$ denote the reconstructed signal predicted by our method and the measured signal strength data, respectively. The loss focusing on the reconstruction of the RF scene is given by:
\begin{equation*} 
\mathcal{L} = | \mathcal{S}^2 - RSSI_{Rx}|^2.
\end{equation*}
During the RF rendering stage, the RF-aware attributes 
$\{\alpha, R, |\boldsymbol{\Gamma}|, \angle \boldsymbol{\Gamma}\}$ 
are optimized through inverse rendering of the observed RF signals.


\section{Implementation and Evaluation} 
We evaluate the proposed RFIR framework from three perspectives: (1) synthesizing RCS for various objects; (2) predicting RSSI in an indoor classroom environment; and (3) demonstrating the framework’s ability to synthesize dynamic wireless environments through scene reconfiguration. Additional training and implementation details are provided in Appendix~\ref{App:Implementation Details}, and the code for reproducing our results is included in the supplementary material.

\begin{figure}
    \centering
    \includegraphics[width=1\linewidth]{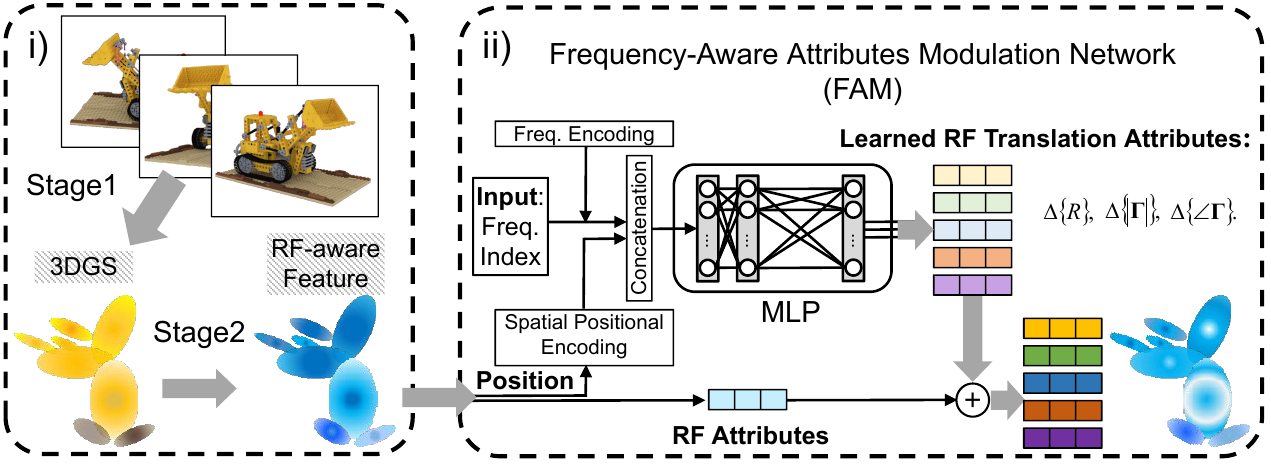}
    \caption{Frequency-Aware Attribute Modulation Network (FAM). We present a novel approach that employs attribute control together with a deformation MLP to modulate the frequency-dependent behavior of 3D Gaussians. The transformed primitives are then rendered into broadband radio signals.}
    \label{fig:deformation}
\end{figure}

\subsection{Case Study I: RCS Synthesis}
RCS characterizes how objects reflect electromagnetic waves and is crucial for understanding signal propagation and designing reliable wireless communication systems. We evaluate the framework's RCS synthesis performance across diverse objects, view points, frequencies, and distances. To broaden our research horizons and ensure versatility, we employ a two-pronged approach combining numerical simulations with real-world experiments.  

\paragraph{Experimental Settings.} The numerical synthetic datasets are generated using Blender in combination with the Mitsuba~3 rendering engine~\cite{han2025rayloc}. Our measured experiments adopt the same monostatic geometric configuration as the simulations, as shown in Fig.~\ref{fig:LFM_setup}. Further details on data acquisition and construction are provided in Appendix~\ref{App:ImplementationDetailsofLFM}.

\paragraph{Baseline Methods.} To validate our RFIR framework, we conduct a comparative analysis against three leading baselines, evaluating RCS synthesis across different frequencies, distances, and object geometries. The baselines include traditional ray tracing, NeRF$^2$~\cite{zhao2023nerf2}, and WRF-GS~\cite{wen2025wrf}. Further details on these methods are provided in Appendix~\ref{App:baseline}.

\paragraph{Architecture Customisation and Wideband Deformation Network.} The 3DGS model is initially trained to reconstruct the scene’s geometry and appearance from multi-view RGB images. It is then trained to predict RCS across diverse observation viewpoints using RF inverse rendering.


To enable wideband RCS synthesis, we incorporate a frequency deformation network into the RFIR framework. As shown in Fig.~\ref{fig:deformation}, this network takes frequency as input and fine-tunes the pre-trained frequency-dependent RF parameters $\{R_j, |\boldsymbol{\Gamma}_{j}|, \angle \boldsymbol{\Gamma}_{j}\}$ for wideband RCS. The deformation network is implemented as a 6-layer MLP with 256 hidden units. This deformable mechanism allows the model to effectively adapt to different frequency bands~\cite{yang2025gsrf,huang2024sc}.

\begin{figure}
    \centering
    \includegraphics[width=1\linewidth]{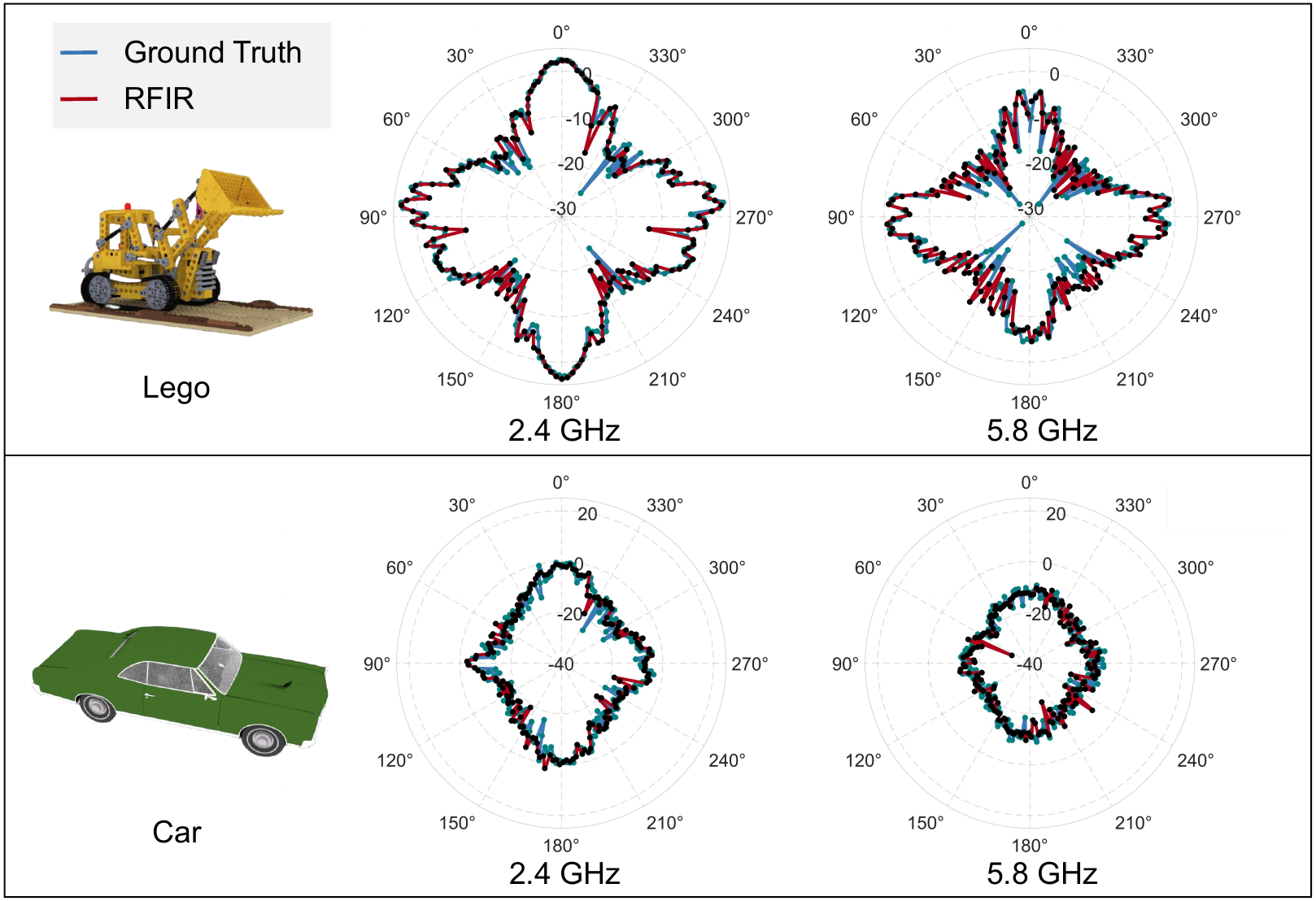}
    \caption{Predicted RCS of the Lego and car at 2.4 and 5.8 GHz.}
    \label{fig:car_lego}
\end{figure}

\paragraph{RCS Synthesis Results.} The results are shown in Fig.~\ref{fig:car_lego}. The predicted errors of RCS at the frequency of 2.4 GHz and 5.8 GHz are shown in Table~\ref{tab:RCS_comparison}. Our method achieves mean absolute error (MAE) of 1.87 dB at 2.4 GHz and 2.69 dB at 5.8 GHz for the Lego model, significantly outperforming NeRF$^2$ and WRF-GS. The superior performance demonstrates the effectiveness of the RF-BSDF parameterization in capturing complex scattering characteristics. 


\begin{figure}[t]
    \centering
    \includegraphics[width=1\linewidth]{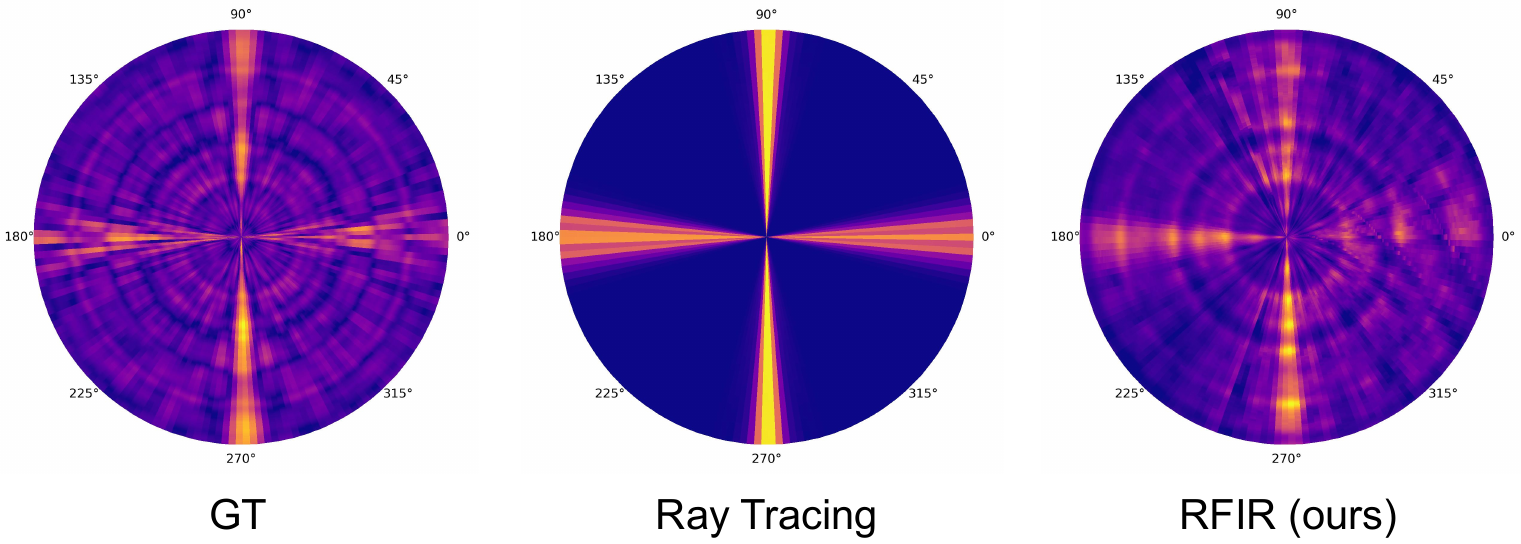}
    \caption{The reconstructed wideband RCS results. }
    \label{Fig:LFM_results}
\end{figure}

\begin{table}[t]
\centering
\caption{Comparison of methods for RCS prediction.}
\label{tab:RCS_comparison}
\resizebox{\columnwidth}{!}{
\begin{tabular}{l|cc|cc}
\toprule
\multirow{2}{*}{Method}       & \multicolumn{2}{c|}{Lego}  & \multicolumn{2}{c}{Car}           \\
             & 2.4 GHz        & 5.8 GHz                    & 2.4 GHz           & 5.8 GHz        \\ \midrule
NeRF$^2$     & 4.82           & 5.23                       & 4.08              & 2.51            \\
WRF-GS       & 2.17           & 2.96                       & 2.54              & 2.33            \\
RFIR (Ours) & \textbf{1.87}  & \textbf{2.69}              & \textbf{2.53}     & \textbf{2.29}  \\ \bottomrule
\end{tabular}
}
\end{table}
We evaluate the efficacy of RFIR by predicting wideband signals from measured linear frequency modulated (LFM) data. To handle the 160 MHz bandwidth consisting of 2000 discrete frequency samples, we extend our single-frequency model (trained at 5.8 GHz) to the wideband domain. This is accomplished via a deformable RF attributes MLP, which adaptively modulates the Gaussian attributes to capture frequency-dependent variations. The prediction results are shown in Fig.~\ref{Fig:LFM_results}. RFIR achieves a low MAE of 1.69 dB, demonstrating high fidelity in wideband spectrum synthesis. In contrast, ray tracing methods yield a much higher MAE of 62.2 dB, indicating their difficulty in accurately modeling complex material properties and surface scattering of physical objects.



\subsection{Case Study II: RSSI Prediction}
\paragraph{Experimental Settings.} RFIR supports highly flexible RSSI prediction. We evaluate the prediction mechanism: generating a radio map for coverage estimation applications. To construct a comprehensive dataset, we employ an open-source classroom model~\cite{BitterliResources16} imported into Blender and rendered using the Mitsuba 3 engine (See Fig.~\ref{fig:classroom} for the schematic and detailed data, provided in Appendix~\ref{App:IRMAP}.).

\paragraph{Baseline Methods.} We compare RFIR with NeRF$^2$, and WRF-GS. We also include FERMI~\cite{luo2025fermi}, which encodes scene information using a geometric map and predicts the signal strength of Tx-Rx pairs at different locations through several neural rendering networks.
\paragraph{Architecture Customization.} We introduce a geometry-conditioned attenuation module that learns a scalar mixing weight over direct and indirect signal components based on Tx-Rx visibility and distance, yielding robust RSSI estimation under arbitrary Tx-Rx configurations. Further design details are provided in Appendix~\ref{App:LoSNLoS}.

\begin{figure}[t]
    \centering
    \includegraphics[width=0.96\linewidth]{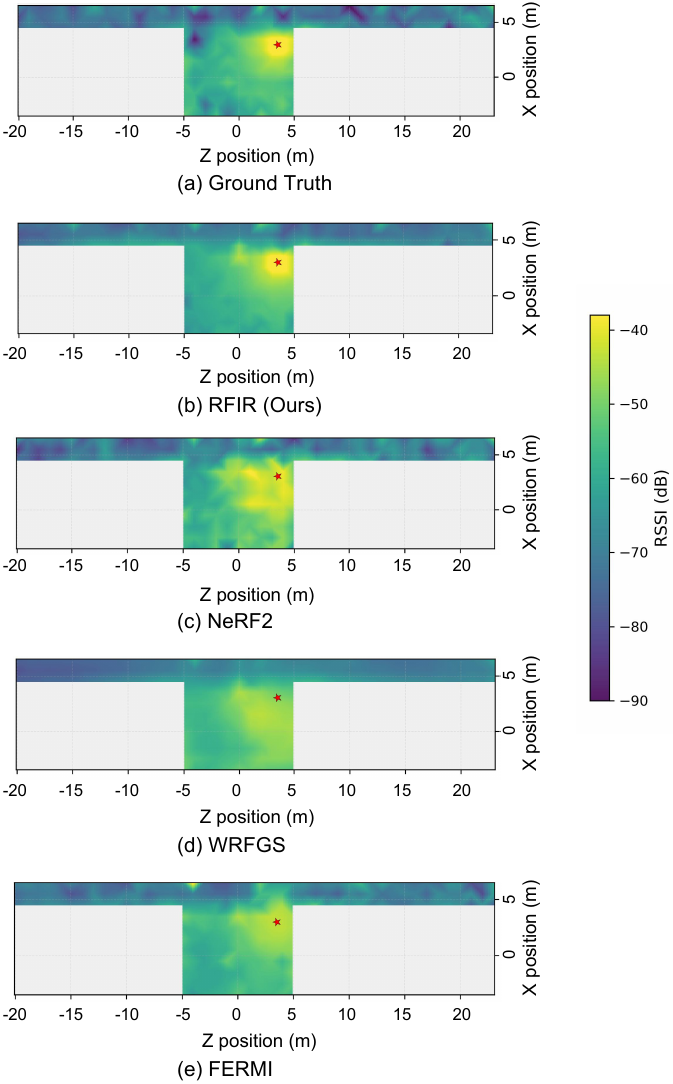}
    \caption{Predicted radio map of the classroom.}
    \label{fig:map}
\end{figure}

\paragraph{RSSI Prediction Results.} 
A comparative assessment of the predicted radio maps is provided in Fig.~\ref{fig:map}. The proposed method achieves an MAE of 4.82 dB, outperforming NeRF2, WRFGS, and FERMI by margins of 0.57 dB, 0.28 dB, and 0.4 dB, respectively. We also present in Appendix~\ref{App:IRMAP} the map visualizations of the three types of RF physical parameters proposed in our inverse rendering method from the antenna’s FoV. This performance gain stems from our proposed decoupling of environment-domain inputs and outputs, which enables the RFIR framework to efficiently characterize wireless scattering in complex environments. In addition, the customized architecture flexibly captures signal propagation under diverse LoS and NLoS conditions, allowing more precise modeling of RF interactions. 

\subsection{Case Study III: System Reconfiguration} 
A reconfigurable wireless system enables dynamic optimization of wireless environments via signal redistribution and environment reconfiguration. RFIR can satisfy the requirements of diverse operational scenarios without the need for exhaustive re-measurement. Experiments focus on signal redistribution to test the RFIR's generalization capabilities. 

\begin{figure}[t]
    \centering
    \includegraphics[width=0.55\linewidth]{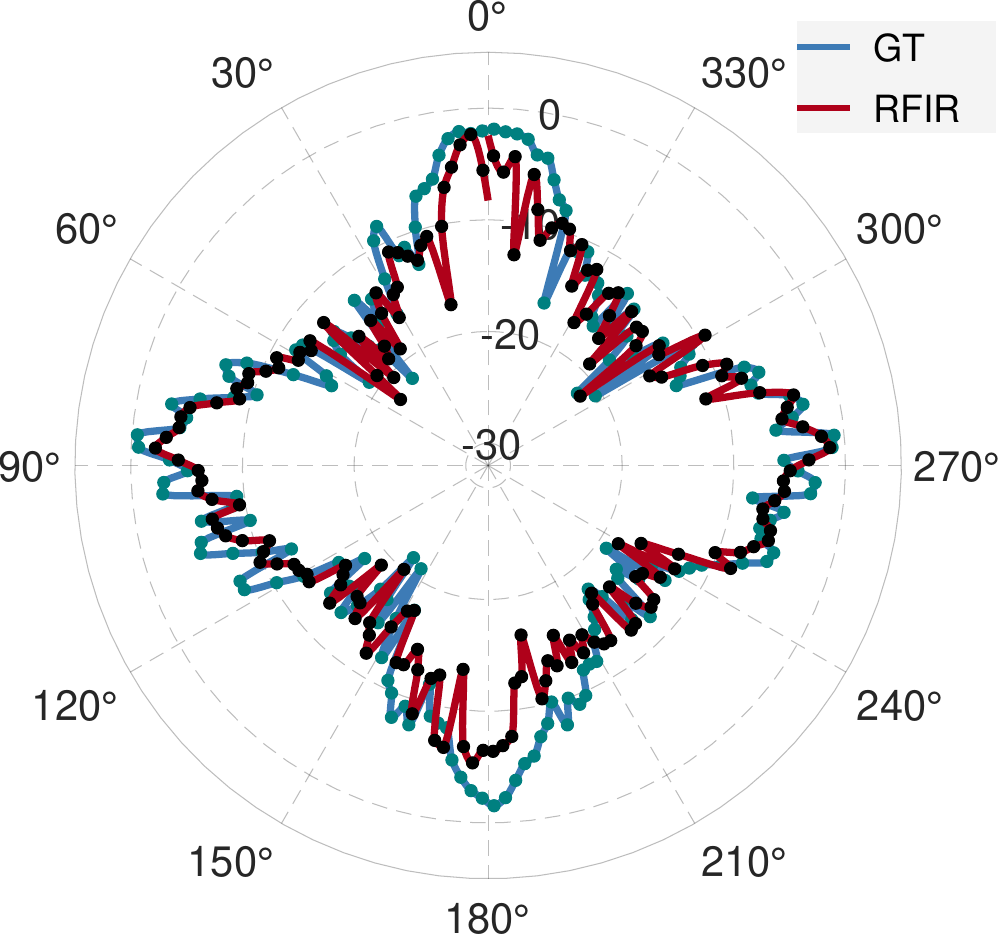}
    \caption{Predicted RCS of the target at 3 m and 2.4 GHz.}
    \label{fig:Edited24GRCS}
\end{figure}

\paragraph{Spatial RCS Extrapolation.} Characterizing the RCS across varying distances is traditionally an arduous and resource-intensive process. We demonstrate that by training on observations at a fixed nominal range, our framework can accurately extrapolate and predict the RCS at multiple disparate distances, significantly reducing measurement overhead. We train the model using data at 2 m, 2.5 m, 4 m, 4.5 m, and 5 m, and test on unseen data at 3 m. As shown in Fig.~\ref{fig:Edited24GRCS}, the proposed method achieves an MAE of 2.62 dB at 2.4 GHz for the Lego model, demonstrating its generalization to intermediate distances.


\subsection{Ablation Study}

\paragraph{Fine-grained Signal Synthesis.} Our full model explicitly incorporates distance-dependent path loss and phase shifts along the Tx–Gaussian–Rx propagation paths. Removing these propagation-aware terms while retaining only BSDF-based scattering leads to a significant degradation in RCS synthesis. In particular, the reconstructed RCS values for the Lego model drop to 3.87 and 6.69, and those for the car model decrease to 2.67 and 2.55. These results highlight the necessity of fine-grained path loss and phase modeling for accurate RF signal synthesis.

\paragraph{LoS/NLoS Weighting Strategy.} We evaluate the effect of balancing LoS and NLoS propagation by directly summing their contributions without dedicated weighting network. This simplification reduces RSSI prediction accuracy by $\sim$35\%. The performance drop indicates that, in complex indoor environments, balancing LoS and NLoS components is essential for balancing direct and reflected energy contributions and improving RF reconstruction quality.

%

\section{Conclusion}
We present RFIR, a physically grounded RF inverse rendering framework that explicitly decouples RF emission, scene geometry, and material electromagnetic properties. By embedding an RF-aware BSDF into the 3D Gaussian splatting paradigm and modeling fine-grained RF propagation via ray tracing, RFIR enables efficient and physically consistent RF signal synthesis and inversion. The proposed decomposition generalizes naturally across multiple RF tasks, while consistently outperforming existing RF neural rendering methods. Our results demonstrate that physically decoupled RF representations provide a powerful foundation for accurate, interpretable, and flexible wireless world modeling.


\section*{Impact Statement}
This paper presents work whose goal is to advance the field of Machine Learning. There are many potential societal consequences of our work, none which we feel must be specifically highlighted here.

\bibliography{RFIR_References}
\bibliographystyle{icml2026}
\newpage
\appendix

\onecolumn

\section{Computing Projected Cross-section of Gaussians}\label{App:projectedcrosssection}
For each Gaussian ellipsoid representing a scattering primitive, the projected cross-section along the transmitter direction $\mathbf{n}_i$ is computed as
\begin{equation*}
\mathcal{A}(\mathbf{n}_i) = \pi \sqrt{\det(\Sigma_i)} \cdot \sqrt{\mathbf{n}_i^\top \Sigma_i^{-1} \mathbf{n}_i},
\end{equation*}
where $\Sigma_i \in \mathbb{R}^{3 \times 3}$ encodes the shape and orientation of the $i$-th ellipsoid, and $\mathbf{n}_i \in \mathbb{R}^3$ is the unit vector from the transmitter to the ellipsoid. The first term, $\pi \sqrt{\det(\Sigma_i)}$, captures the overall scale of the ellipsoid, while the second term, $\sqrt{\mathbf{n}_i^\top \Sigma_i^{-1} \mathbf{n}_i}$, adjusts the cross-section according to the projection direction. This projected area determines the fraction of incident energy intercepted by the scattering primitive, effectively weighting its contribution in the RFIR rendering process.

\section{Antenna FoV Discretization}\label{App:Discretization}
Different FoV discretization schemes are employed depending on the antenna type. For omnidirectional antennas, the virtual reception aperture is discretized over an elevation of $180^\circ$ and azimuth of $360^\circ$ at $1^\circ$ intervals. For directional horn antennas, the reception surface is discretized over $90^\circ$ elevation and $360^\circ$ azimuth at $1^\circ$ intervals.

\section{Training and Implementation Details}\label{App:Implementation Details}
As described in Sec.~\ref{sec:System Design}, the training procedure consists of two stages. In the first stage, we initialize 100,000 Gaussians per object and optimize the model for 30,000 iterations using the Adam optimizer. The initial learning rate is set to 0.01 and gradually decayed to 0.001. In the second stage, the RFIR parameters are introduced and jointly optimized with the Gaussian representation for various downstream tasks. To enable GPU-accelerated RFIR training and feature extraction in 3D Gaussian splatting, we implement the BSDF parameter evaluation in CUDA. Furthermore, a BVH-based visibility computation of the incident field is used to efficiently handle large-scale ray interactions.

\section{Implementation Details for Experiments on Data Acquisition.}~\label{App:ImplementationDetailsofLFM}
Specifically, 3D models of various objects~\cite{BitterliResources16}, including a car and a Lego assembly, are used to capture multi-view RGB images along with the corresponding RF signal data, simulated via ray tracing at 2.4 GHz and 5.8 GHz. 
The distance between the target object and the co-located transceiver is varied from 2 m to 5 m, and each object is placed on a rotating platform to systematically acquire RCS measurements across all aspect angles. 
Half of the data are used for training, with the remainder reserved for testing, ensuring uniform coverage across rotation angles.

For real-world experiments, multi-view images of the target objects are captured using a standard camera, while wideband RCS is measured simultaneously. 
Measurements are performed with a USRP X310 software-defined radio platform equipped with a pair of horn antennas, capable of operating across a 160 MHz to 5.8 GHz frequency range. 
A metallic target in the shape of the letter ``H'' with physical dimensions $0.5\,\mathrm{m} \times 0.2\,\mathrm{m} \times 0.6\,\mathrm{m}$ is used as the object under test. 
The target is mounted on a rotating turntable with an angular resolution better than $0.0001^\circ$. 
At every $1^\circ$ rotation interval, both RGB images and the corresponding scattered wireless signal strength are collected at a distance of 3~m. 
To ensure accurate camera geometry, intrinsic calibration and precise extrinsic parameters are performed. 
For wireless signal generation, a wideband LFM signal with a 160~MHz bandwidth is uniformly discretized into 2000 frequency samples in the frequency domain.

\section{Geometric Optimization}\label{App:Geometric Regularization}
\subsection{Geometric Regularization.} To ensure structural fidelity, we implement a multi-objective geometric regularization strategy.We enforce depth-normal consistency by aligning the rendered normal map $\mathcal{N}$ with a pseudo-normal $\tilde{\mathcal{N}}$ derived from the depth gradient $\nabla \mathcal{D}$ under a local planarity assumption. The corresponding consistency loss is formulated as:
\begin{equation*} \label{Eq:depth_normal_loss}
    \mathcal{L}_n = |\mathcal{N} - \tilde{\mathcal{N}}|_2.
\end{equation*}
Furthermore, to suppress artifacts and promote spatial smoothness, we introduce a depth uncertainty loss $\mathcal{L}_u$ and an edge-aware normal smoothness loss $\mathcal{L}_{s,n}$:
\begin{equation*} \label{Eq:D_loss}
    \mathcal{L}_u = \mathcal{D}_{sq} - \mathcal{D}^2, \quad \mathcal{L}_{s,n} = |\nabla \mathcal{N}| \exp(-|\nabla C_{\text{gt}}|),
\end{equation*}
where $\mathcal{D}_{sq}$ represents the second moment of the depth distribution (see Appendix~\ref{App:Depth Uncertainty} for detailed derivations). While $\mathcal{L}_u$ minimizes the variance along rays to produce sharper surfaces, $\mathcal{L}_{s,n}$ preserves geometric discontinuities by weighting the normal gradient against the intensity edges of the ground-truth image $C_{\text{gt}}$.

\subsection{Depth Uncertainty.}~\label{App:Depth Uncertainty} The term $\mathcal{L}_u = \mathcal{D}_{sq} - \mathcal{D}^2$ measures the variance of the depth of Gaussians that contribute to a specific pixel. Mathematically, it is expressed as:
\begin{equation*}
\mathcal{L}_u = \sum_{i=1}^N \omega_i d_i^2 - \left( \sum_{i=1}^N \omega_i d_i \right)^2,
\end{equation*}
where $\omega_i = \alpha_i T_i$. In wireless channel modeling, multi-path components are highly sensitive to surface precision. Minimizing this uncertainty penalizes ``cloud-like'' or semi-transparent Gaussian clusters, forcing the primitives to concentrate on the actual physical boundary of objects.

To summarize, the first stage of our framework represents a 3D scene as a set of enhanced geometry Gaussian primitives, where the $i$-th Gaussian $\mathcal{P}_i$ is parameterized as $\{\boldsymbol{\mu}_i, \Sigma_i, \alpha_i,\boldsymbol{n}_i\}$.

\subsection{Loss Function Design}~\label{App:Loss Function}
In the geometric reconstruction phase, we optimize the 3DGS framework by incorporating a suite of regularizers alongside standard photometric losses. Specifically, we integrate depth-normal consistency, depth distribution constraints, and normal smoothness regularizers. The total objective function for this stage is defined as:
\begin{equation*} 
\mathcal{L}=\lambda_{1} \mathcal{L}_{1}+\lambda_{s s i m} \mathcal{L}_{s s i m}+\lambda_{n} \mathcal{L}_{n} + \lambda_{u} \mathcal{L}_{u} + \lambda_{s, n} \mathcal{L}_{s, n},
\end{equation*}
where $\lambda_{1}, \lambda_{s s i m}, \lambda_{n}, \lambda_{s, n}, \lambda_{u}$ represent the weighting coefficients for the pixel-wise photometric loss, SSIM loss, depth-normal consistency, depth uncertainty and normal smoothness, respectively.

\section{Baselines of RCS synthesis}\label{App:baseline}
\paragraph{Ray Tracing.} For numerical evaluation, a ray tracing framework integrated with a BSDF model is employed to generate high-fidelity RCS ground truth. In measured experiments, this framework, utilising meshes reconstructed from captured imagery, serves as a theoretical baseline. The simulation models electromagnetic propagation via a transmitter-target-receiver path, explicitly incorporating Fresnel reflections according to geometric optics principles.
\paragraph{NeRF$^2$~\cite{zhao2023nerf2}.} We adopt the NeRF$^2$ framework as a volumetric learning-based baseline for RCS synthesis. NeRF$^2$ employs a neural radiance network to model the radiance field, enabling the synthesis of wireless signals from arbitrary viewpoints. 
\paragraph{WRF-GS~\cite{wen2025wrf}.} WRF-GS is a recent work that extends 3DGS for wireless signal synthesis. We model the RCS using 3D Gaussians as the secondary radiator, WRF-GS synthesizes the received signal at the receiver through volumetric rendering techniques.

\section{Further Information and Results on RSSI Prediction}\label{App:IRMAP}
\begin{figure}
    \centering
    \includegraphics[width=0.8\linewidth]{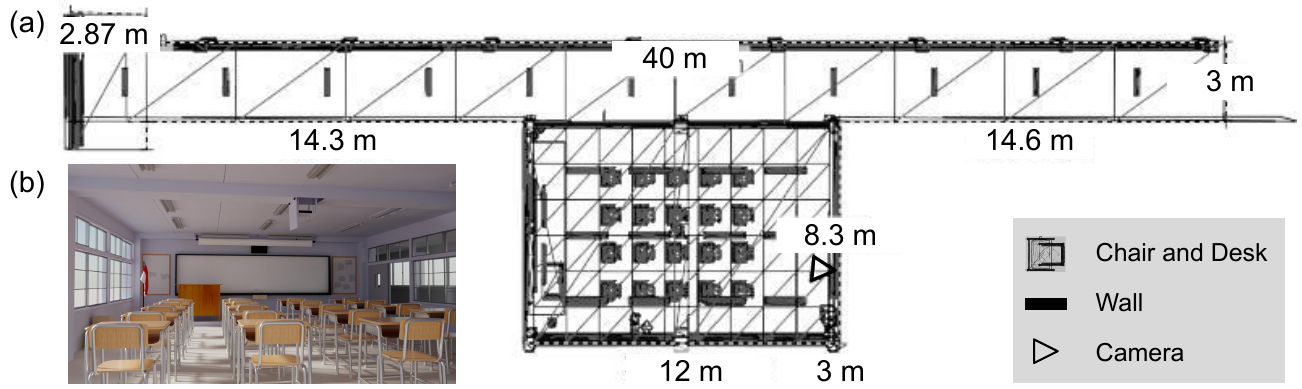}
    \caption{3D classroom models for simulation purposes. We provide (a) the classroom dimensions along with (b) an image captured by the camera.}
    \label{fig:classroom}
\end{figure}

\paragraph{Data Acquisition.} A schematic of the 3D classroom models is shown in Fig.~\ref{fig:classroom}. By configuring various camera poses, we generate multi-view images to facilitate 3DGS reconstruction of the classroom environment. 
For the radio map generation task, omnidirectional Tx antennas are deployed at 24 distinct locations, and receiving antennas collect RSSI data from 440 sampling points per Tx location. The dataset is then randomly split into an 80\% training set and a 20\% testing set.

\begin{figure}
    \centering
    \includegraphics[width=1\linewidth]{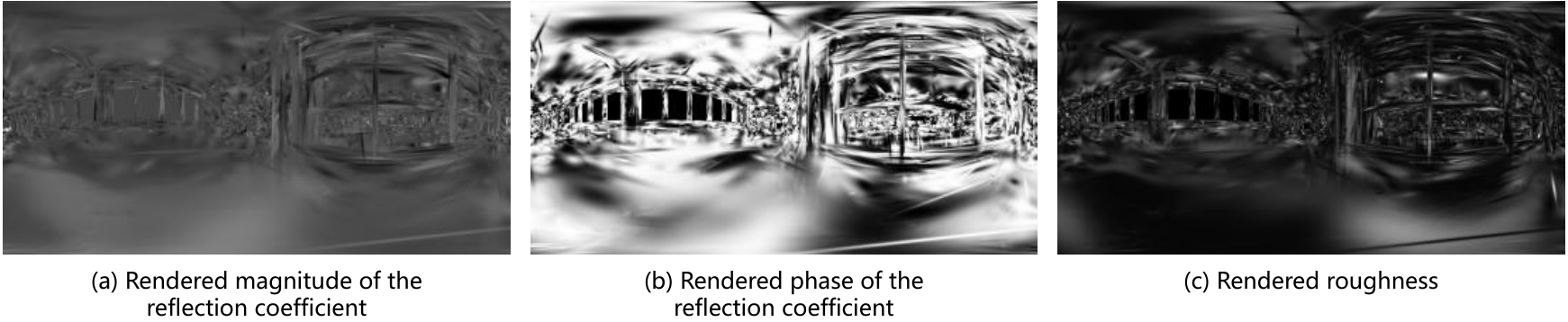}
    \caption{Visualization of the rendered (a) magnitude and (b) phase of the reflection coefficient, as well as the rendered (c) roughness maps, obtained via inverse rendering of the classroom scene.}
    \label{fig:inverse_rendering}
\end{figure}

\paragraph{More results.} More results on our inverse rendering are presented as shown in Fig.~\ref{fig:inverse_rendering}. 

\section{Physics-Guided LoS/NLoS Signal Mixing.}\label{App:LoSNLoS}

We decompose the LoS/NLoS signal synthesis into a direct LoS component and a scattered NLoS component rendered by RFIR. Specifically, the signal propagation process is modeled in two stages, where the dominant LoS component is first estimated, followed by the modeling of the NLoS component via RFIR rendering.

\paragraph{LoS Modeling.}
A distance-aware network is employed to estimate the initial transmit signal strength $S_{\text{Tx}}$ and the spatial attenuation coefficient $c_{\text{dis}}$, providing a coarse yet physically meaningful characterization of the LoS component.

\paragraph{NLoS Modeling.}
With the LoS parameters fixed, the NLoS component rendered via RFIR is coherently combined with the LoS signal. We introduce a visibility term $v_{\text{vis}} \in [0,1]$ to represent the probability of an unobstructed direct path between the Tx and Rx. 

Accordingly, the received signal can be expressed as $ \hat{\mathcal{S}}(\mathbf{x}_t, \mathbf{x}_r) = \mathcal{S}_{\text{LoS}} + \mathcal{S}_{\text{NLoS}}$, where the LoS component is defined as $\mathcal{S}_{\text{LoS}} = v_{\text{vis}} \cdot S_{\text{Tx}} \cdot c_{\text{dis}} \cdot e^{j \phi_{\text{dis}}}$, with the phase shift $\phi_{\text{dis}} = \frac{2\pi d_{\text{LoS}}}{\lambda}$, where $d_{\text{LoS}}$ denotes the propagation distance and $\lambda$ is the wavelength. The NLoS component $\mathcal{S}_{\text{NLoS}}$ is estimated via the RFIR rendering equation.

This formulation preserves the physical inductive bias of wireless signal propagation, while enabling end-to-end optimization of transmit power, attenuation coefficients, and NLoS scattering parameters, resulting in a flexible and accurate model for complex propagation environments.


\end{document}